\documentclass[twocolumn,aps,prc,superscriptaddress,nofootinbib,showpacs,floatfix,longbibliography]{revtex4}
\usepackage{url}
\usepackage{cancel}
\usepackage[colorlinks,linkcolor=blue,citecolor=blue,filecolor=black,urlcolor=blue]{hyperref}
\usepackage{epsfig,graphics}
\usepackage{graphicx}
\usepackage{dcolumn}
\usepackage{bm}
\usepackage{amssymb}
\usepackage{amsmath}
\usepackage{multirow}
\usepackage{float}
\usepackage{harpoon}
\usepackage{MnSymbol}
\usepackage{appendix}
\usepackage{color}
\usepackage{hyperref}
\usepackage{cleveref}

\newcommand{\nch} {N_{\mathrm{ch}}}
\newcommand{\snn}{\mbox{$\sqrt{s_{\mathrm{NN}}}$}}
\newcommand{\pT} {p_{\mathrm{T}}}
\newcommand{\lr}[1]{\left\langle #1\right\rangle}

\newcommand{\ruru}{$^{96}$Ru+$^{96}$Ru}
\newcommand{\zrzr}{$^{96}$Zr+$^{96}$Zr}

\begin{document}
\title{Energy dependence of heavy-ion initial condition in isobar collisions }
\newcommand{\bnl}{Physics Department, Brookhaven National Laboratory, Upton, NY 11976, USA}
\newcommand{\sbu}{Department of Chemistry, Stony Brook University, Stony Brook, NY 11794, USA}
\author{Somadutta Bhatta}\affiliation{\sbu}
\author{Chunjian Zhang}\affiliation{\sbu}
\author{Jiangyong Jia}\email{jiangyong.jia@stonybrook.edu}\affiliation{\sbu}\affiliation{\bnl}
\date{\today}
\begin{abstract}
Collisions of isobar nuclei, those with the same mass number but different structure parameters, provide a new way to probe the initial condition of the heavy ion collisions. Using transport model simulation of $^{96}$Ru+$^{96}$Ru and $^{96}$Zr+$^{96}$Zr collisions at two energies $\sqrt{s_{\mathrm{NN}}}=0.2$ TeV and 5.02 TeV, where $^{96}$Ru and $^{96}$Zr nuclei have significantly different deformations and radial profiles, we identify sources of eccentricities contributing independently to the final state harmonic flow $v_n$. The efficacy for flow generation differs among these sources, and explains the modest energy dependence of the isobar ratios of $v_n$. Additionally, a significant component of $v_n$ is found to be uncorrelated with the eccentricity but is instead generated dynamically during system evolution. Experimental measurement of these ratios at the LHC energy and comparison with RHIC energy can provide insight into the collision-energy dependence of the initial condition. 
\end{abstract}
\pacs{25.75.Gz, 25.75.Ld, 25.75.-1}
\maketitle

\textit{\bf Introduction.} 
One major challenge in heavy ion phenomenology is the need to simultaneously determine the ``initial condition'' and the ``transport properties'' of the quark-gluon plasma (QGP), each of which has multiple parameters. Current state-of-the-art multi-system Bayesian analyses show that the parameters of the initial condition and the transport properties are correlated, such that the simultaneous extraction of both is required to improve the precision of the extracted QGP properties~\cite{Bernhard:2016tnd, Everett:2020xug, Nijs:2020ors}. The connection between initial condition and final state observables is relatively well understood within a given model, for example, $v_n\propto \varepsilon_n$ for $n=2$ and 3~\cite{Teaney:2010vd}. However, the initial condition can not be calculated directly from the first principle QCD theory. Thus, it is desirable to identify experimental observables that can more directly pinpoint signatures of the initial condition.

One promising approach is to consider the collision of isobar systems with the same mass number but different yet well-known structural properties. These properties can be characterized, for instance, by parameters of a deformed Woods-Saxon distribution with varying nuclear shape and radial profile. In experimental measurements, we focus on the ratio of a given observable $\mathcal{O}$ in collisions of isobars $X$ and $Y$, and ask $\mathcal{O}_{\rm X+X}/\mathcal{O}_{\rm Y+Y } \stackrel{?}{=} 1$. Any significant departure from unity must originate from the structural differences, which impacts the initial condition and survives to the final state~\cite{Giacalone:2021uhj}. Studies have been performed for ratios of bulk observables such as the distribution of charged particle multiplicity $p(\nch)$, harmonic flow $v_2$ and $v_3$ and mean transverse momentum $\lr{\pT}$~\cite{Li:2019kkh,Xu:2021uar,Jia:2021oyt,Nijs:2021kvn}. These ratios are found to be insensitive to the final state effects~\cite{Zhang:2022fou} and hence reflect mainly how the initial condition responds to variations in the nuclear structure. 

At high energy, the initial condition for bulk particle production is dominated by partons at a small longitudinal momentum fraction $x$. The distribution of these partons depends on not only distributions of nucleons from nuclear structure input, but also modifications due to gluon shadowing or gluon saturation effects, encapsulated in the so-called nuclear parton distribution function (nPDF)~\cite{Rojo:2019uip, AbdulKhalek:2022fyi}. The nPDF and, hence, the emergent initial condition is naturally expected to vary with $\snn$ and pseudorapidity $\eta$. For example, the initial condition at mid-rapidity is controlled by low-$x$ partons from both nuclei. In contrast, at forward rapidity, it is controlled by large-$x$ partons from the projectile nucleus and small-$x$ partons from the target nucleus. The small-$x$ evolution may smooth out the long-range density variation induced by nuclear structure in a $\snn$ and rapidity-dependent manner~\cite{Albacete:2014fwa, Helenius:2016dsk}. Hence, comparing isobar ratios at different $\snn$ and $\eta$ can be helpful to detect these novel nPDF effects~\cite{Bally:2022vgo}.

In this paper, we perform a first study of $\snn$ and $\eta$ dependence the isobar ratios in $^{96}$Ru+$^{96}$Ru and $^{96}$Zr+$^{96}$Zr collisions for several bulk observables, $p(\nch)$, $v_2$ and $v_3$. Our study is based on a popular transport model (AMPT)~\cite{Lin:2004en}, which previously were shown to be able to reproduce the experimental isobar ratios measured by the STAR Collaboration~\cite{Xu:2021uar, Jia:2021oyt, Nijs:2021kvn}.

\textit{\bf Setup.} The AMPT data at $\snn=200$ GeV were produced in a previous study~\cite{Jia:2022qgl}. Therefore, we only need to generate the isobar data at $\snn=5.02$ TeV using AMPT version 2.29t9. The model was run in the string-melting mode with a partonic cross-section of 3.0~mb~\cite{Ma:2016fve}. The nucleon distribution in colliding ions is parameterized by a deformed Woods-Saxon distribution,
\begin{align}\label{eq:1}
\rho(r,\theta,\phi)&\propto\frac{1}{1+e^{[r-R_0\left(1+\beta_2 Y_2^0(\theta,\phi) +\beta_3 Y_3^0(\theta,\phi)\right)]/a_0}},
\end{align}
which includes four parameters: quadrupole deformation, $\beta_2$, octupole deformation, $\beta_3$, half-width radius, $R_0$, and surface diffuseness, $a_0$. It was already established those isobar ratios are controlled by parameter differences, $\Delta \beta_2^2 = \beta_{\mathrm{2Ru}}^2-\beta_{\mathrm{2Zr}}^2$, $\Delta \beta_3^2 = \beta_{\mathrm{3Ru}}^2-\beta_{\mathrm{3Zr}}^2$, $\Delta a_0 =a_{\mathrm{0Ru}}-a_{\mathrm{0Zr}}$ and $\Delta R_0 =R_{\mathrm{0Ru}}-R_{\mathrm{0Zr}}$~\cite{Zhang:2021kxj}. Therefore, we simulate generic isobar $^{96}$X+$^{96}$X collisions with five choices of $\beta_2$, $\beta_3$, $R_0$ and $a_0$ listed in Table~\ref{tab:1}. This allows us to define ratios that isolate influences of the nuclear structure parameters step-by-step. The $\nch$, $v_2$ and $v_3$ are calculated using particles within $0.2<\pT<2$ GeV/$c$ in various $\eta$ ranges. The effects of non-flow are evaluated by comparing results obtained from the default standard method, which uses all unique pairs, and the two-subevent method, where the particles in pairs are taken from the opposite $\eta$ range~\cite{Jia:2017hbm}. The small difference of $v_n$ between the two methods is consistent with the impact of longitudinal flow decorrelations~\cite{Nie:2022gbg}. The impact of non-flow, which has a characteristic $1/\nch$ in $v_n$, is visible only at $\nch<50$ in the AMPT model in the standard method.

\begin{table}[!h]
\centering
\caption{\label{tab:1} Nuclear structure parameters used in the simulations of \ruru{} and \zrzr{} collisions. Case1 and Case5 represent, respectively, full parameterizations of $^{96}$Ru and $^{96}$Zr.} 
\begin{tabular}{|l|cccc|}\hline 
   &\; $R_0$ (fm)\; & \;$a_{0}$ (fm)\;  & $\beta_{2}$ & $\beta_{3}$  \\\hline 
Case1 $^{96}$Ru & 5.09  & 0.46   & 0.162 & 0  \\
Case2          & 5.09  & 0.46   & 0.06  & 0  \\
Case3          & 5.09  & 0.46   & 0.06 & 0.20  \\
Case4          & 5.09  & 0.52   & 0.06 & 0.20  \\
Case5 $^{96}$Zr & 5.02  & 0.52   & 0.06 & 0.20  \\\hline
\end{tabular}
\end{table}

The initial condition in the AMPT model is controlled by the HIJING model, where the particle production is described in terms of a soft and a hard component~\cite{Lin:2004en}. The soft component produces particles in a string picture, which scales as the mass number $A$ and increases slowly with $\snn$. In contrast, the hard component produces particles via pQCD minijet partons described by the impact parameter dependent nPDF, which scales as $A^{4/3}$ and grows fast with $\snn$. In our simulation, particle production at $\snn=0.2$ TeV is dominated by the soft component, whereas the hard component is greatly enhanced at $\snn=5.02$ TeV. Note that the nPDF implemented in HIJING is quite old, and captures only partially $\snn$ and $\eta$ dependence of the initial condition. 

\begin{figure}[!h]
\includegraphics[width=1\linewidth]{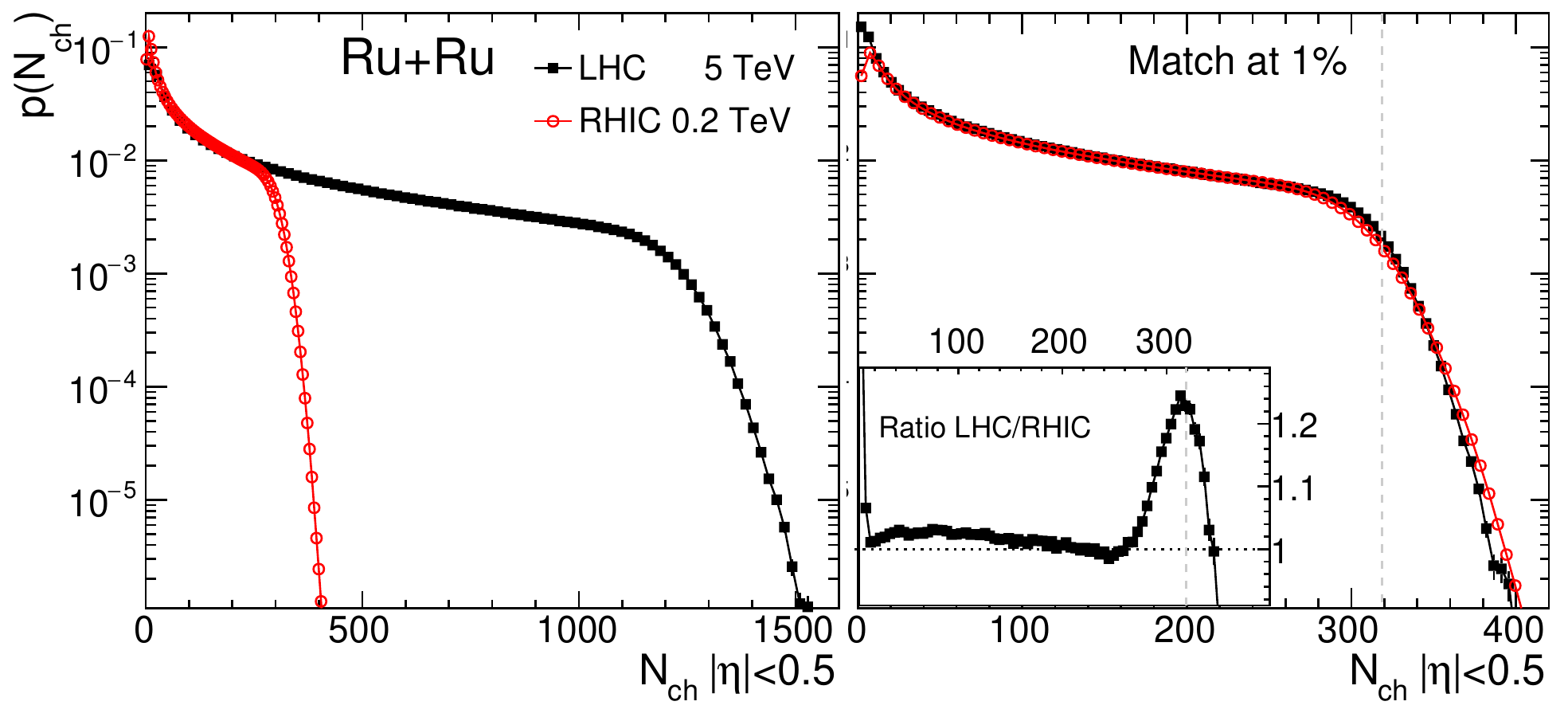}
\caption{\label{fig:1} The distributions of $\nch$, $p(\nch)$, in Ru+Ru collisions at $\snn=0.2$ TeV and 5 TeV before (left) and after (right) rescaling to match the at 1\% centrality. The configuration of ``Case 1'' in Tab~\ref{tab:1} is used. The insert panel shows the ratio of the rescaled $p(\nch)$.}
\end{figure}

\textit{\bf Results.} Figure~\ref{fig:1} displays the distribution of charged particle multiplicity $p(\nch)$ at mid-rapidity $|\eta|<0.5$. The range in $\nch$ is much larger at the LHC energy. However, after applying a scale factor of 1/3.9, which matches the $\nch$ values at 1\% centrality at the two energies, the two distributions have very similar shapes. The shape of $p(\nch)$ at the LHC is sharper, due to a somewhat better centrality resolution. Their difference in shape, quantified by the ratio in the insert panel, reveals a shallow bump in mid-central collisions and a sharp bump in central collisions. However, these differences are mild considering the large increase in $\nch$ from RHIC to the LHC. For a fair study of the $\snn$ dependence, the results from LHC are presented as a function of the scaled $\nch$ match to the value at RHIC energy. 

\begin{figure}[!h]
\includegraphics[width=0.9\linewidth]{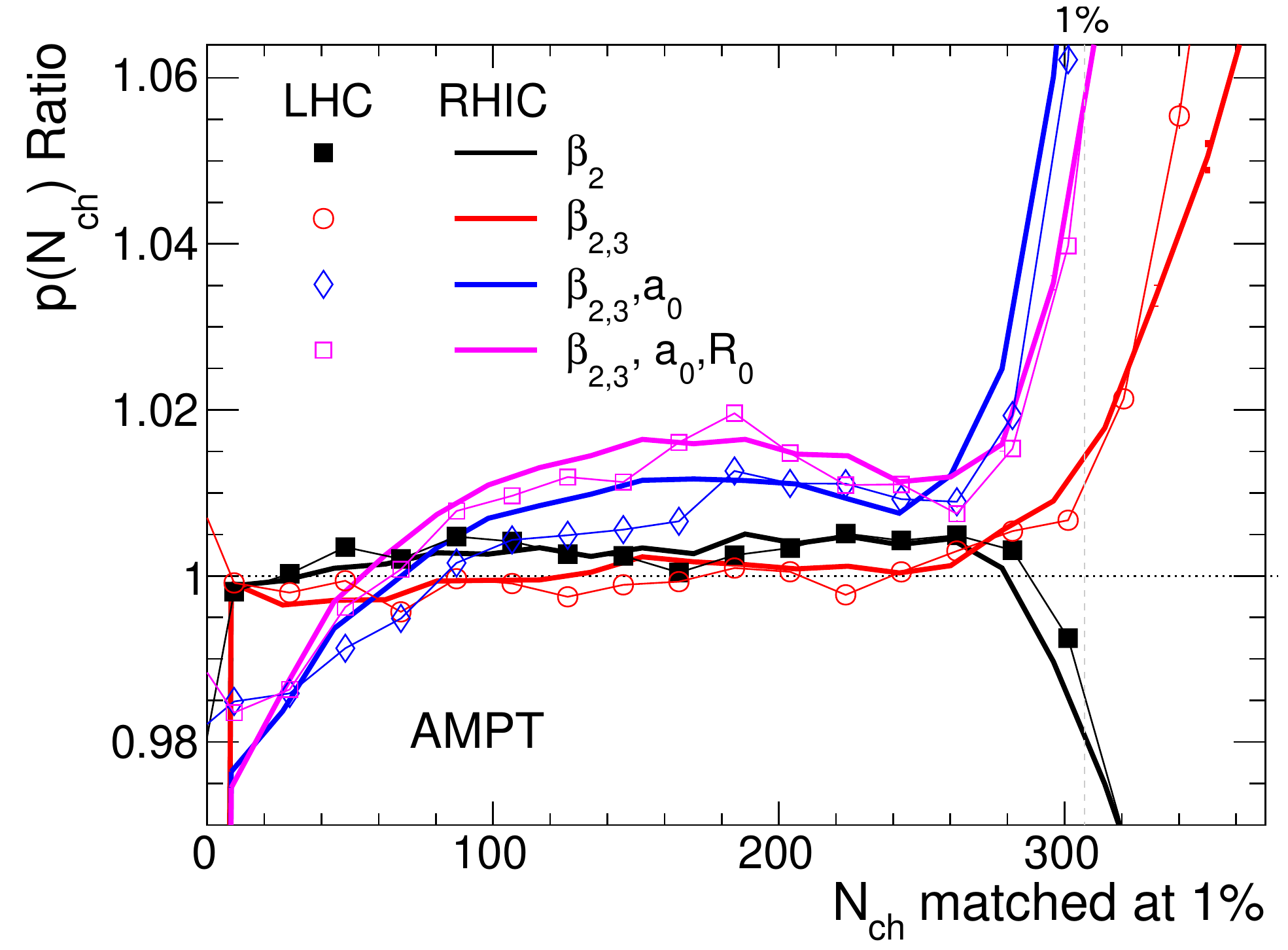}
\caption{\label{fig:2} The ratios of $p(\nch)$ including the differences of the four nuclear structure parameters $\beta_2$, $\beta_3$, $a_0$ and $R_0$, step-by-step: $\frac{\mathrm{\textstyle\small Case1}}{\mathrm{ \textstyle\small Case2}}$\;,\; $\frac{\mathrm{\textstyle\small Case1}}{\mathrm{ \textstyle\small Case3}}$\;,\;$\frac{\mathrm{\textstyle\small Case1}}{\mathrm{ \textstyle\small Case4}}$\;,\;$\frac{\mathrm{\textstyle\small Case1}}{\mathrm{ \textstyle\small Case5}}$. The results at RHIC energy $\snn=0.2$ TeV are shown by solid lines, while those at the LHC energy $\snn=5$ TeV are shown by symbols. }
\end{figure}

Figure~\ref{fig:2} shows the four isobar ratios of $p(\nch)$ using the setting in Table~\ref{tab:1}, calculated separately at RHIC and the LHC, which include the effects of nuclear structure step by step. These ratios have been studied in detail in Ref.~\cite{Jia:2021oyt} and were found to describe the STAR data. We found that the ratios at the LHC energy show nearly the same behavior. Such a lack of $\snn$ dependence suggests that the ratios of $p(\nch)$ are a robust probe for structure difference between isobar systems. 

\begin{figure}[!t]
\includegraphics[width=0.8\linewidth]{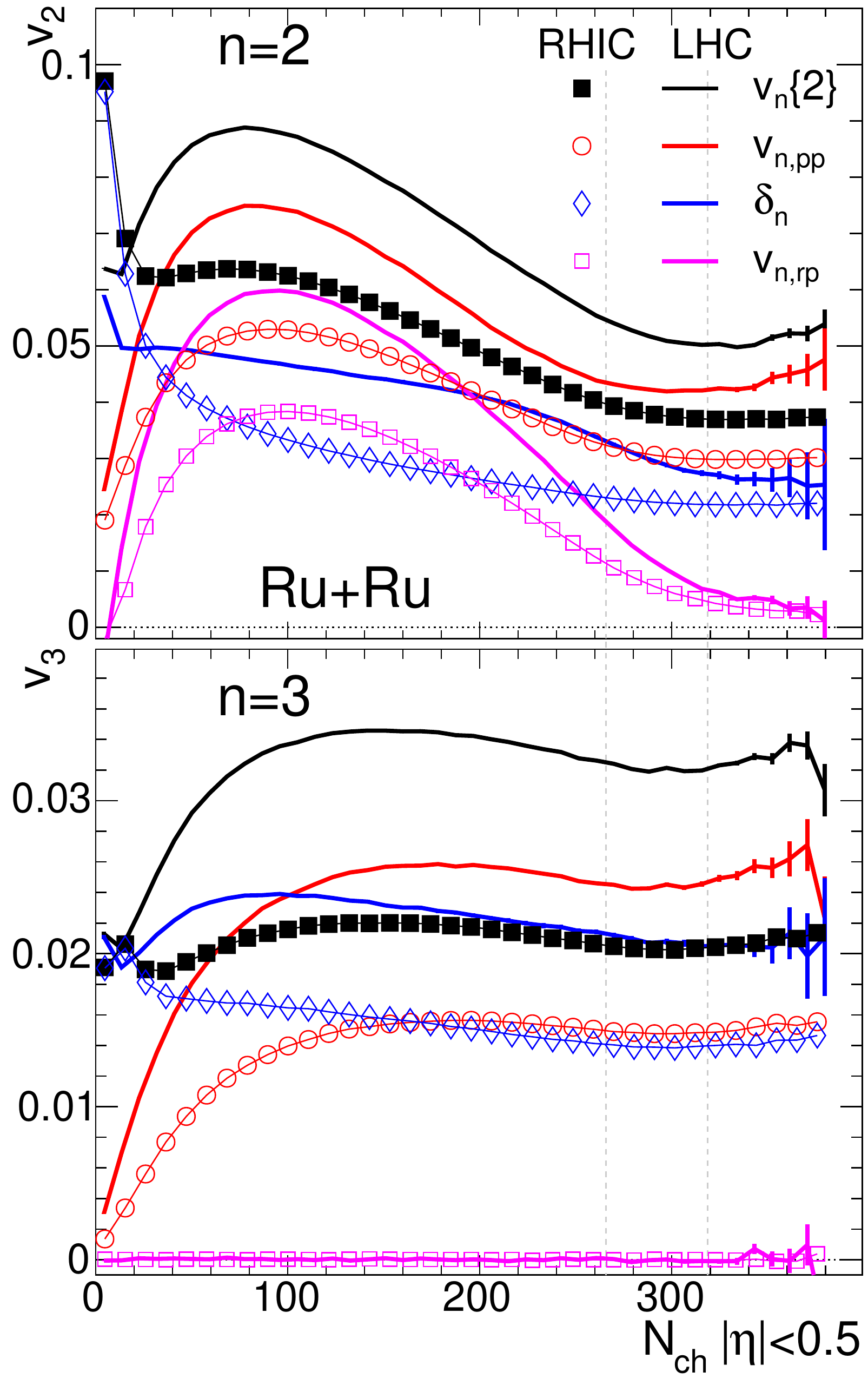}
\caption{\label{fig:3} The results of harmonic flow in Ru+Ru collisions, calculated via two-particle correlation $v_n\{2\}$, participant plane $v_{n,\mathrm{pp}}$, the difference $\delta_n = \sqrt{(v_n\{2\})^2-(v_{n,\mathrm{pp}})^2}$, and the reaction plane $v_{n,\mathrm{rp}}$ from RHIC energy in symbols and LHC energy in solid lines for $n=2$ (left panel) and $n=3$ (right panel). The LHC $\nch$ was scaled by 1/3.9 to match $p(\nch)$ at RHIC.}
\end{figure}

Next, we study the energy dependence of harmonic flow. Figure~\ref{fig:3} compares the flow calculated using the two-particle correlation method $v_{n}\{2\}$ between the two energies. The $v_{n}\{2\}$ values are significantly larger at the LHC, which is expected as stronger collective flow is generated from more frequent partonic scattering. An increase of the $v_n\{2\}$ values at $\nch<50$, was observed at RHIC but not LHC, implying that non-flow contribution for a given centrality is smaller at higher energy.

To understand the $v_n\{2\}$ in terms of initial collision geometry, we calculated the harmonic flow vector $V_n=v_ne^{in\Psi_n}$ with respect to the ``participant plane'' (PP) eccentricity  $\mathcal{E}_n = \varepsilon_ne^{in\Phi_n}$,
\begin{align}\label{eq:2}
v_{n,\mathrm{pp}} = \frac{|\left\langle V_n\mathcal{E}_n^*\right\rangle|}{\sqrt{\left\langle \mathcal{E}_n\mathcal{E}_n^*\right\rangle}}\;,
\end{align}
which captures the flow generated by the collision geometry. Another related quantity is the flow with respect to the impact parameter direction, $v_{n,\mathrm{rp}}$, also known as the flow with respect to the reaction plane (RP), which quantifies the strength of flow driven by the average geometry of the overlap region. Both $v_{n,\mathrm{pp}}$ and $v_{n,\mathrm{rp}}$ are smaller than $v_n\{2\}$, as observed in Fig.~\ref{fig:3}. As expected, the $v_{2,\mathrm{rp}}$ reaches zero in central collisions, and $v_{3,\mathrm{rp}}$ vanishes everywhere since there is no average triangular component in the collision geometry.

The difference between $v_n\{2\}$ and $v_{n,\mathrm{pp}}$ suggests that flow from the two-particle correlation method has a component largely uncorrelated with initial collision geometry, which we define as
\begin{align}\label{eq:3}
\delta_n = \sqrt{(v_n\{2\})^2-(v_{n,\mathrm{pp}})^2}\;.
\end{align}
The $\delta_n$ are also shown in Fig.~\ref{fig:3}, whose magnitude is quite comparable to $v_{n,\mathrm{pp}}$. Namely, the value of $\delta_2$ is only slightly smaller than  $v_{2,\mathrm{pp}}$ in central collisions, and the values of $\delta_3$ are comparable to $v_{3,\mathrm{pp}}$ in mid-central and central collisions, and are larger in peripheral collisions. The values of $\delta_n$ have a weak dependence on $\nch$, suggesting most of them are unrelated to the non-flow since the latter is expected to have a $1/\nch$ dependence. Instead, $\delta_n$ could originate from the hydrodynamic responses to local hot spots in the initial state, which has been shown to increase the flow fluctuations but are uncorrelated with the participant eccentricity~\cite{Pang:2015zrq,Jia:2024xvl}.

\begin{figure}[!h]
\includegraphics[width=0.8\linewidth]{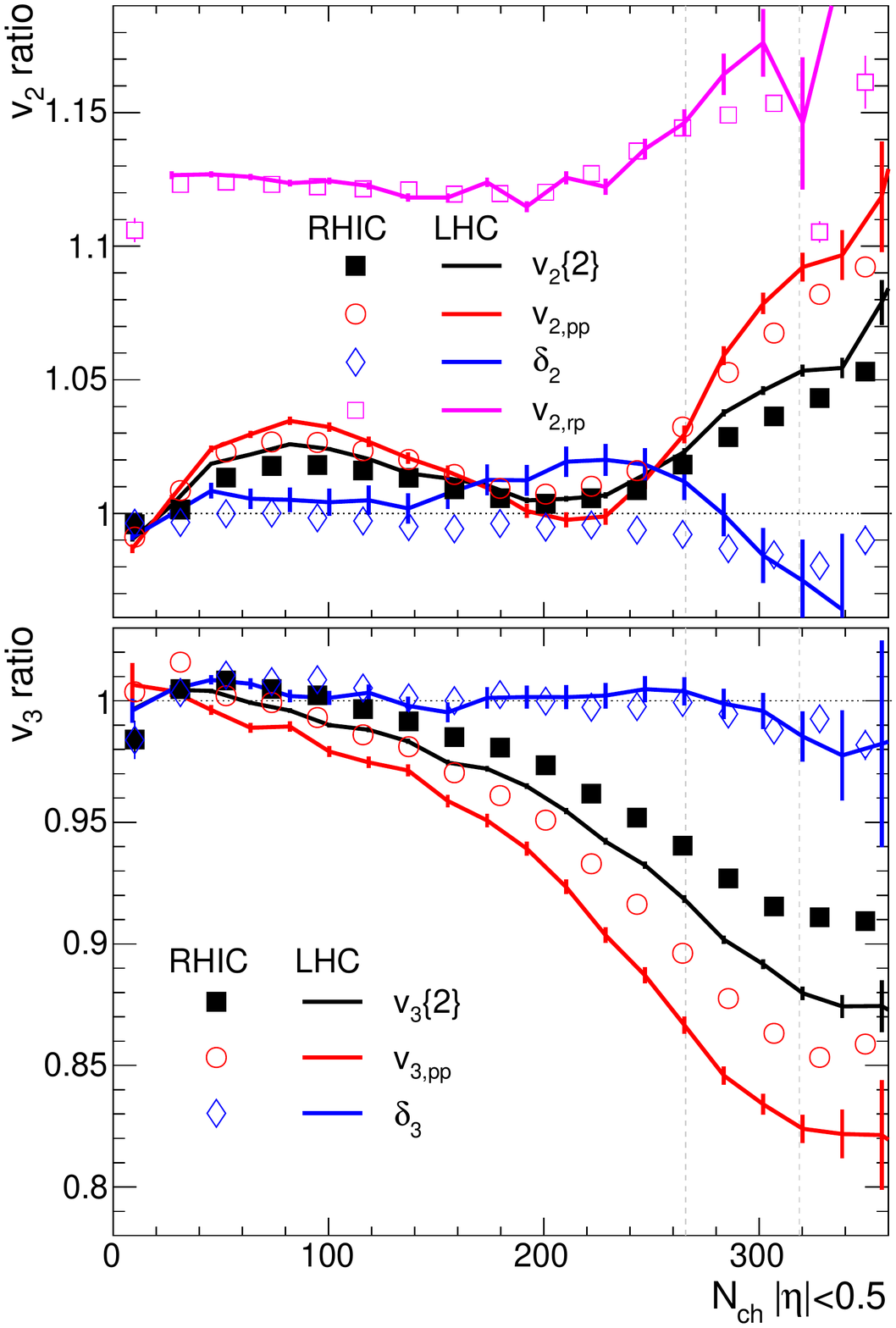}
\caption{\label{fig:4} The isobar ratios (Case1/Case5 in Table~\ref{tab:1}) for $v_n\{2\}$, $v_{n,\mathrm{pp}}$, their differences $\delta_n\equiv \sqrt{(v_n\{2\})^2-(v_{n,\mathrm{pp}})^2}$, and $v_{n,\mathrm{rp}}$ for $n=2$ (top panel) and $n=3$ (bottom panel), calculated at RHIC energy $\snn=0.2$ TeV in symbols and LHC energy $\snn=5$ TeV in solid lines.  The LHC $\nch$ was scaled by 1/3.9 to match $p(\nch)$ at RHIC.}
\end{figure}

Previous studies have shown that the ratios of $v_n\{2\}$ are sensitive to nuclear structure parameter differences between isobar or isobar-like systems~\cite{Jia:2021oyt, Nijs:2021kvn, Jia:2022qgl,STAR:2024eky}. Here, we want to investigate how such sensitivity shows up individually for various components of the harmonic flow: $v_{n,\mathrm{pp}}$, $\delta_n$, and $v_{n,\mathrm{rp}}$. The results are shown in Fig.~\ref{fig:4}, where we show the ratio Case1/Case5 in Table~\ref{tab:1}, which includes the full nuclear structure differences between Ru and Zr. One striking feature is that the ratios of $\delta_n$ are very close to unity, whereas the ratios of $v_{n,\mathrm{pp}}$ have a much larger deviation from unity than the ratio of $v_n\{2\}$. This behavior implies that $\delta_n$ is insensitive to variations of nuclear structure parameters, and hence simply dilutes the nuclear structure dependence of $v_n\{2\}$. As mentioned earlier, $\delta_n$ may be caused by local correlations in the initial state or arise dynamically as a stochastic source in the final state. In the future, it would be interesting to repeat this study in smaller systems for which experimental data already exist, such as $p$/$d$/$^{3}$He+Au collisions~\cite{ATLAS:2012cix, PHENIX:2018lia,STAR:2022pfn,STAR:2023wmd}; where $v_{n}\{2\}$ could be dominated by $\delta_n$, and the much smaller $v_{n,\mathrm{pp}}$ component should reflect better the ordering of the $\varepsilon_n$ between these systems. Lastly, the ratio of $v_{n,\mathrm{rp}}$ shows a large enhancement from unity, which was clarified in Ref.~\cite{Jia:2022qgl} as being dominated by the $a_0$ difference between Ru and Zr. 

\begin{figure}[!h]
\includegraphics[width=1\linewidth]{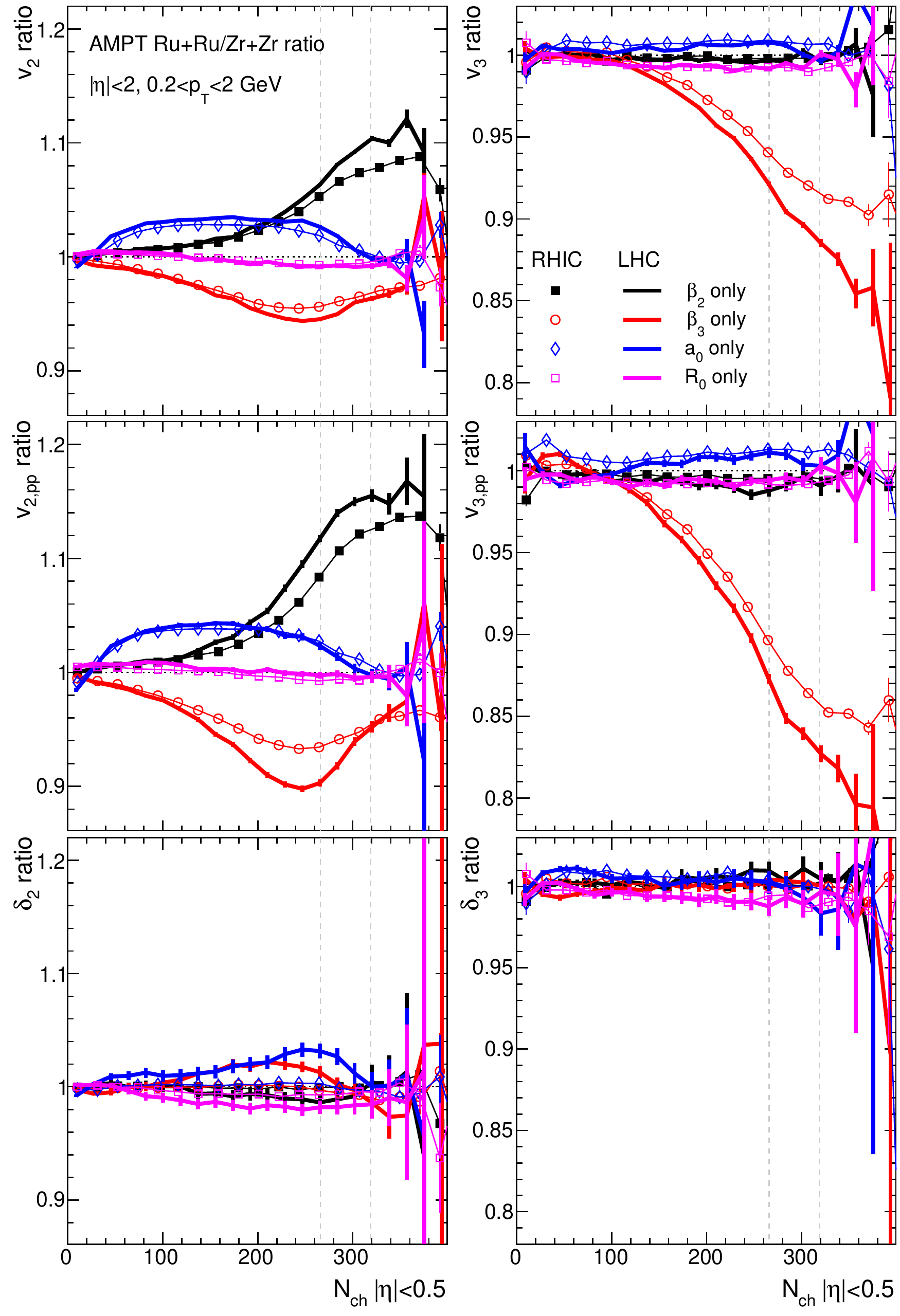}
\caption{\label{fig:5}  The isobar ratios for $v_n\{2\}$ (top row), $v_{n,\mathrm{pp}}$ (middle row), and their differences $\delta_n\equiv \sqrt{(v_n\{2\})^2-(v_{n,\mathrm{pp}})^2}$ (Bottom row) for $n=2$ (left column) and $n=3$ (right column), calculated at RHIC energy $\snn=0.2$ TeV in solid lines and LHC energy $\snn=5$ TeV in solid symbols. In each panel, the ratios are shown separately for the impact of $\beta_2$, $\beta_3$, $a_0$, and $R_0$ between Ru+Ru and Zr+Zr collisions.  The LHC $\nch$ was scaled by 1/3.9 to match $p(\nch)$ at RHIC.}
\end{figure}

It is well known that the impact of all four structure parameters, $\beta_2$, $\beta_3$, $a_0$, and $R_0$, depend on the observable and centrality. Therefore, it is important to study their influences separately. The results are shown in Fig.~\ref{fig:5}. For all four parameters, the ratios of $v_{n,\mathrm{pp}}$ show stronger variations than the ratios of $v_n\{2\}$, and the ratios of $\delta_n$ are always much closer to unity than the ratios of $v_n\{2\}$ or $v_{n,\mathrm{pp}}$.

Regarding energy dependence, the behavior of the ratios in Figs.~\ref{fig:4} and \ref{fig:5} are qualitatively similar between RHIC and the LHC energies. However, we notice that the deviations from unity are systematically more significant at the LHC than at RHIC, which is particularly evident for $\beta_2$ and $\beta_3$. These behaviors can be understood from the energy dependence of the flow response coefficients, which we shall elaborate as follows.

In general, the mean square eccentricity may have several independent sources, $\lr{\varepsilon_n^2} = \sum_i \lr{\varepsilon_{n;i}^2}$. Correspondingly, the final state harmonic flow would also have multiple sources, $v_n\{2\}^2 = \delta_n^2+ \sum_i k_{n;i}^2 \lr{\varepsilon_{n;i}^2}$, where $\delta_n$ is the flow uncorrelated with initial eccentricity, and $k_{n;i}$ are the response coefficients. Our main point is that the values of response coefficients and their $\snn$ dependencies are not the same between different sources in the AMPT models. We shall demonstrate this point using the elliptic flow as an example.

In the presence of nuclear deformation, the mean-square elliptic eccentricity and elliptic flow can be expressed as,
\begin{align}\nonumber 
\varepsilon_2^2 &= \varepsilon_{2,0}^2+ \varepsilon_{2,{\mathrm{rp}}}^2+a\beta_2^2+b\beta_3^2\;,\\\label{eq:4}
v_2\{2\}^2 &= \delta_2^2 + k_{0}^2\varepsilon_{2;0}^2+ k_{1}^2\varepsilon_{2,{\mathrm{rp}}}^2+k_{2}^2a\beta_2^2+k_{3}^2b\beta_3^2\;,
\end{align}
which contains a component driven by the reaction plane eccentricity $\varepsilon_{2,{\mathrm{rp}}}$, a component arising from fluctuation $\varepsilon_{2,0}$, as well as two comparatively smaller components driven by the quadruple and octupole deformations~\footnote{Eccentricity vector has a $x$ component along the impact parameter and a $y$ component. Formally, our definition, in the absence of deformation, implies $\varepsilon_{2,{\mathrm{rp}}}\equiv \lr{\varepsilon_{2,x}}$ and $\varepsilon_{2,0}^2\equiv \lr{\varepsilon_{2,y}^2} + \lr{(\varepsilon_{2,x}-\lr{\varepsilon_{2,x}})^2}$.}. For conciseness, we drop the ``$\lr{}$'' and the harmonic number in the notation for response coefficients.  Since these components are independent, the participant plane elliptic flow in Eq.~\eqref{eq:2} becomes~\footnote{Since $V_n = k_n V_n\mathcal{E}_n$, the power of the response coefficients should be same as that for $v_n$, i.e. Eq.~\eqref{eq:6} (Eq.~\eqref{eq:4}) has linear (quadratic) dependencies.} 
\begin{align}\label{eq:6}
v_{2,\mathrm{pp}}&= \frac{k_{0}\varepsilon_{2;0}^2+ k_{1}\varepsilon_{2,{\mathrm{rp}}}^2+k_{2}a\beta_2^2+k_{3}b\beta_3^2}{\sqrt{\varepsilon_{2,0}^2+ \varepsilon_{2,{\mathrm{rp}}}^2+a\beta_2^2+b\beta_3^2}}\;.
\end{align}

This equation shows that the relative contribution of deformation, important in central collisions, depends not only on the $\beta_n$ but also on the $k_n$. In particular, if various $k_n$ increase with $\snn$ at different rates, then the isobar ratios would not be the same between RHIC and the LHC, as observed in Figs.~\ref{fig:4} and \ref{fig:5}. The fact that these ratios have a stronger dependence on the variation of $\beta_2$ and $\beta_3$ at the LHC, implies that $k_2$ and $k_3$ increase more strongly with $\snn$ than $k_0$~\footnote{The values of $k_1$ are not relevant in central collisions, since $\varepsilon_{2,{\mathrm{rp}}}$ approaches zero.}. 

Among various components $\varepsilon_{2,\mathrm{pp}}$ in Eq.~\eqref{eq:4}, $\varepsilon_{2;0}$ and $\varepsilon_{2,{\mathrm{rp}}}$ are by far the most dominant. $\varepsilon_{2;0}$ is important in the central and most peripheral collisions, while $\varepsilon_{2,{\mathrm{rp}}}$ is more important in the mid-central collisions (see Fig.~\ref{fig:3}). In the top panel of Fig.~\ref{fig:4}, we notice that the ratio of $v_{2,\mathrm{rp}}$ shows no difference between RHIC and LHC, despite a much larger deviation from unity than the ratio of $v_{2,\mathrm{pp}}$. To understood this behavior, we point out that $v_{2,\mathrm{rp}}$ has only one source, i.e. $v_{2,\mathrm{rp}} = k_1 \varepsilon_{2,{\mathrm{rp}}}$. Even though $k_1$ changes with $\snn$, this change is expected to cancel in the isobar ratio, $v_{2,\rm Ru}^{\mathrm{rp}}/v_{2,\rm Zr}^{\mathrm{rp}} = \varepsilon_{2,\rm rpRu}/\varepsilon_{2,\rm rpZr}$. Furthermore, the energy dependence of $k_0$ and $k_1$ determines the energy dependence of $v_{2,\mathrm{pp}}$. For example, if $\frac{(k_0)_{\rm LHC}}{(k_0)_{\rm RHIC}}<\frac{(k_1)_{\rm LHC}}{(k_1)_{\rm RHIC}}$, then $\frac{(v_{2,\mathrm{pp}})_{\rm LHC}}{(v_{2,\mathrm{pp}})_{\rm RHIC}} \approx \frac{\varepsilon^2_{2;0}(k_{0})_{\rm LHC}+\varepsilon^2_{2,{\mathrm{rp}}} (k_{1})_{\rm LHC}} {\varepsilon^2_{2;0}(k_{0})_{\rm RHIC}+\varepsilon_{2,{\mathrm{rp}}}^2 (k_{1})_{\rm RHIC}}<\frac{(k_1)_{\rm LHC}}{(k_1)_{\rm RHIC}} = \frac{(v_{2,\mathrm{rp}})_{\rm LHC}}{(v_{2,\mathrm{rp}})_{\rm RHIC}}$. These behaviors are qualitatively supported by results in the top panel of Fig.~\ref{fig:3} (i.e. $\frac{(v_{2,\mathrm{pp}})_{\rm LHC}}{(v_{2,\mathrm{pp}})_{\rm RHIC}}\lesssim\frac{(v_{2,\mathrm{rp}})_{\rm LHC}}{(v_{2,\mathrm{rp}})_{\rm RHIC}}$), which implies that in the AMPT model, the relative contribution from average geometry is a bit larger at the LHC energy than at the RHIC energy. 

\begin{figure}[!h]
\includegraphics[width=1\linewidth]{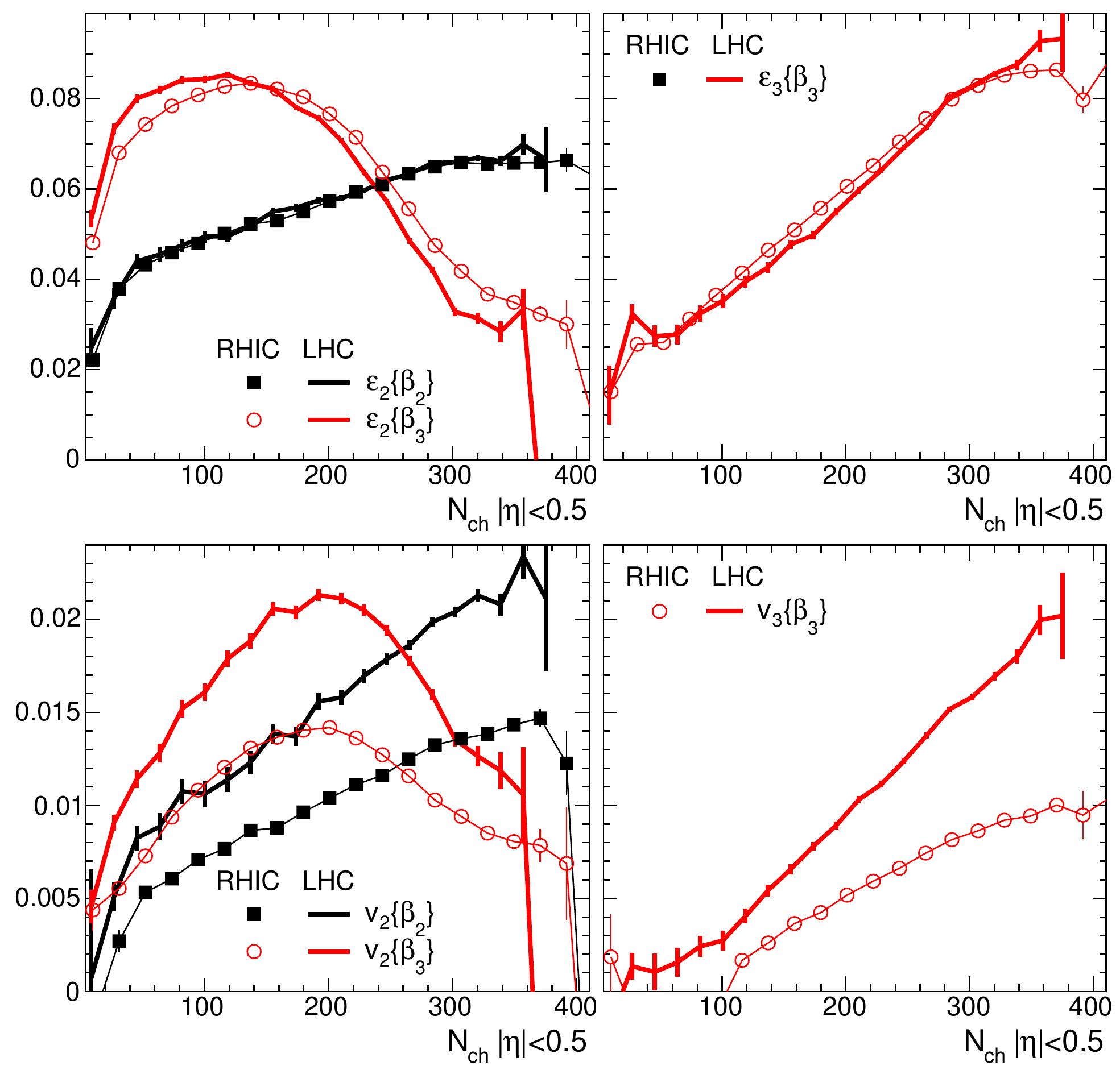}
\caption{\label{fig:6} The contribution from nuclear deformation assuming $\beta_2=0.16$ and $\beta_3=0.2$ to eccentricities, $\varepsilon_2$ (top left) and $\varepsilon_2$ (top right), and flow, $v_2$ (bottom left) and $v_3$ (bottom right) at RHIC energy in symbols and LHC energy in solid lines. They are defined according to Eq.~\eqref{eq:7}.  The LHC $\nch$ was scaled by 1/3.9 to match $p(\nch)$ at RHIC.}
\end{figure}

We then consider the components associated with $\beta_2$ and $\beta_3$ from Eq.~\eqref{eq:4}, which are denoted here by
\begin{align}\nonumber
&\varepsilon_{2}\{\beta_2\}^2 \equiv \varepsilon_{2;2}^2 = a\beta_2^2\;,\; \varepsilon_{2}\{\beta_3\}^2 \equiv \varepsilon_{2;3}^2 = b\beta_3^2\;,\\\label{eq:7}
&v_{2}\{\beta_2\}^2 \equiv  k_{2}^2a\beta_2^2\;,\; v_2\{\beta_3\}^2 = k_{3}^2b\beta_3^2\;.
\end{align} 
Similarly, we can define terms $\varepsilon_{3}\{\beta_3\}$ and $v_3\{\beta_3\}$, accounting for the $\beta_3$ contribution to $\varepsilon_3$ and $v_3$, respectively. These quantities can be calculated by comparing $\varepsilon_{n}^2$ obtained for collisions with and without $\beta_n^2$, i.e. $\varepsilon_{n}\{\beta_n\}^2 = \varepsilon_{n,\rm deformed}^2 - \varepsilon_{n,\rm spherical}^2$. In practice, these quantities are obtained using the realistic values of deformations in the isobar systems $\beta_2=0.16$ and $\beta_3=0.2$. The results obtained from the AMPT simulation are shown in Fig.~\ref{fig:6} at both energies. All three eccentricity observables $\varepsilon_{2}\{\beta_2\}$, $\varepsilon_{2}\{\beta_3\}$, $\varepsilon_{3}\{\beta_3\}$ are nearly the same between RHIC and the LHC, whereas the $\nch$ dependence of $v_{2}\{\beta_2\}$, $v_{2}\{\beta_3\}$, $v_{3}\{\beta_3\}$ are stronger at the LHC energy. Interestingly, the values of $\varepsilon_{n}\{\beta_n\}$ and $v_{n}\{\beta_n\}$ increase nearly linearly with $\nch$.

\begin{figure}[!h]
\includegraphics[width=1\linewidth]{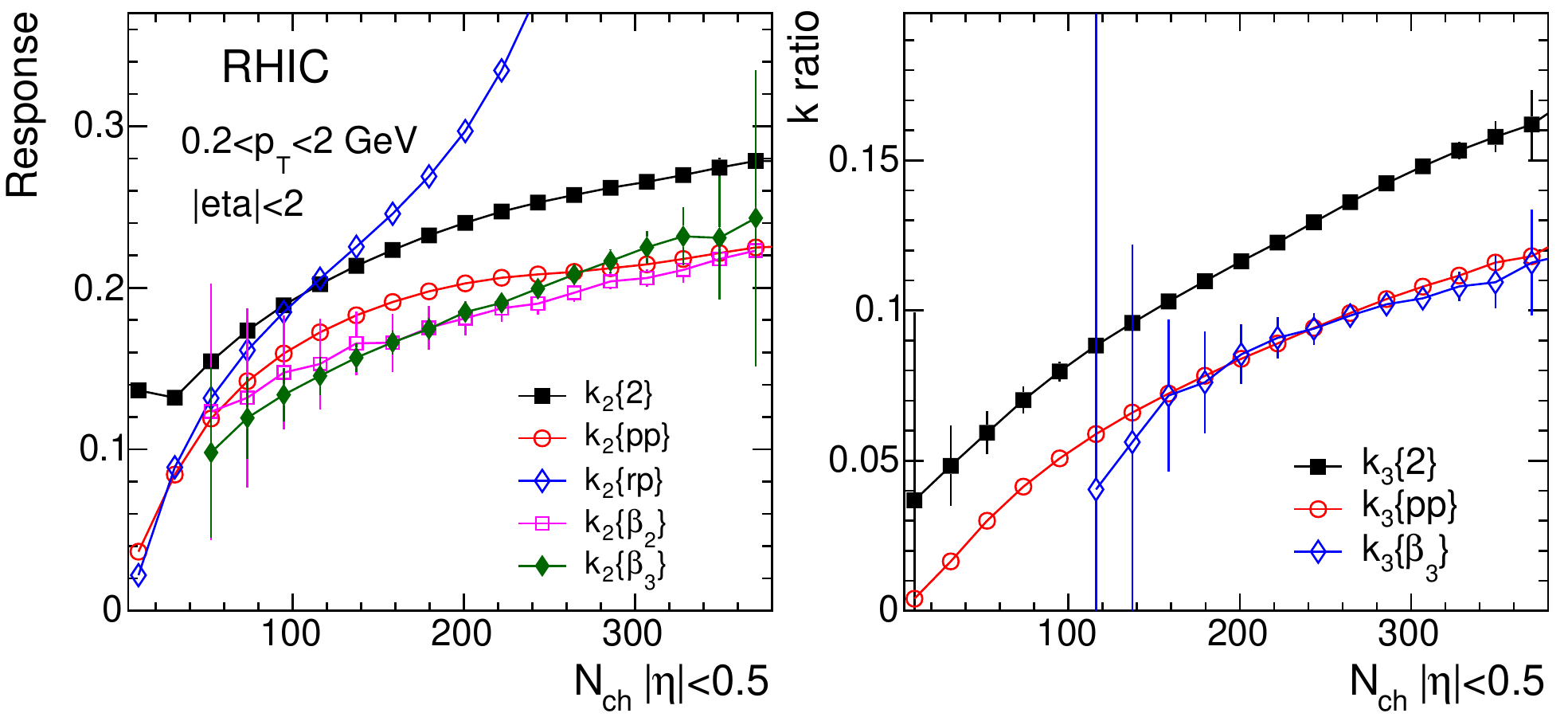}
\caption{\label{fig:6b} The response coefficients for various definition and components of $v_2$ (left) and $v_3$ (right) at the RHIC energy.}
\end{figure}

Figure~\ref{fig:6b} shows the response coefficients for various components of flow at RHIC energy, obtained using the data in Figs.~\ref{fig:3}--\ref{fig:6} and Eq.~\eqref{eq:4}. By construction, $k_n\{2\}\equiv v_n\{2\}/\varepsilon_n$ are greater than $k_n\{\rm pp\}\equiv v_n\{\rm pp\}/\varepsilon_n$. The values of $k_2\{\rm pp\}$ are slightly larger than the values of $k_2\{\beta_n\}$, this enhancement can be naturally attributed to the much larger response of the reaction plane flow $k_2\{\rm rp\}$. On the other hand, the value of $k_3\{\rm pp\}$ are very similar to $k_3\{\beta_3\}$. 

\begin{figure}[!h]
\includegraphics[width=1\linewidth]{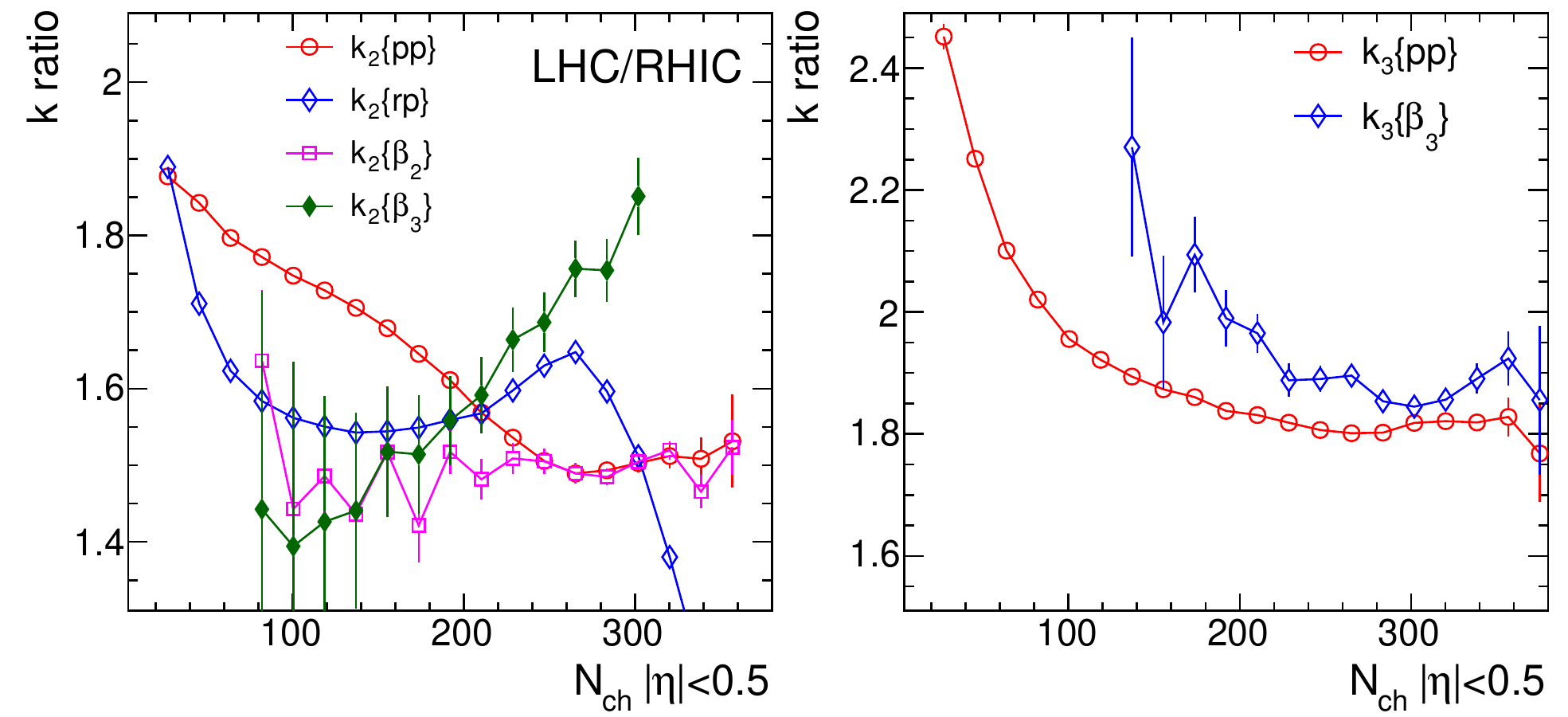}
\caption{\label{fig:7} The ratios of response coefficients between LHC and RHIC for $v_2$ (left) and $v_3$ (right). The LHC $\nch$ was scaled by 1/3.9 to match $p(\nch)$ at RHIC.}
\end{figure}

Next, we quantify the energy dependence of the response of harmonic flow to nuclear deformations. We first calculate the response coefficients at the LHC energy similar to those shown in Fig.~\ref{fig:6b}. We then calculate the ratios of the response coefficients between the LHC and RHIC, and the results are displayed in Fig~\ref{fig:7}. Since eccentricities have very weak energy dependence (Fig.~\ref{fig:6}), these ratios reflect mainly the energy dependence of the $v_n$. The energy dependence of these response coefficients for elliptic flow is different from each other by about 20--30\%: $k_2\{\rm pp\}$ is largest in the mid-central collisions, while toward central collisions, the values of $k_2\{\rm rp\}$ and $k_2\{\beta_3\}$ are larger. In central collisions the values of $k_2\{\beta_2\}$ and $k_2\{\rm pp\}$ approach each other, while the values of $k_3\{\beta_3\}$ are slightly larger than the values of $k_3\{\rm pp\}$. These behaviors are responsible for the stronger impact of nuclear deformation on the isobar ratio at the LHC compared to RHIC, as observed in Figs.~\ref{fig:4} and ~\ref{fig:5}.

\begin{figure}[!h]
\includegraphics[width=1\linewidth]{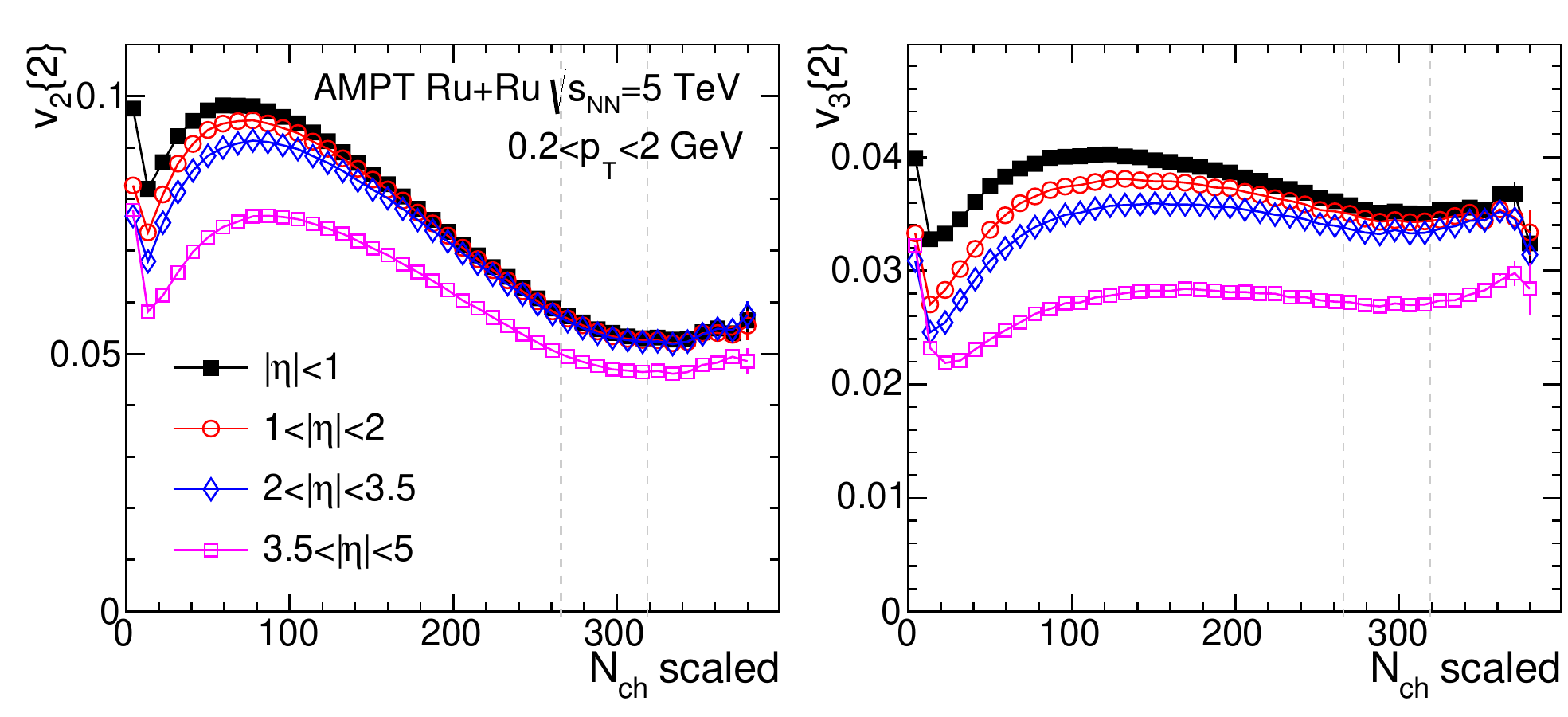}\\
\includegraphics[width=1\linewidth]{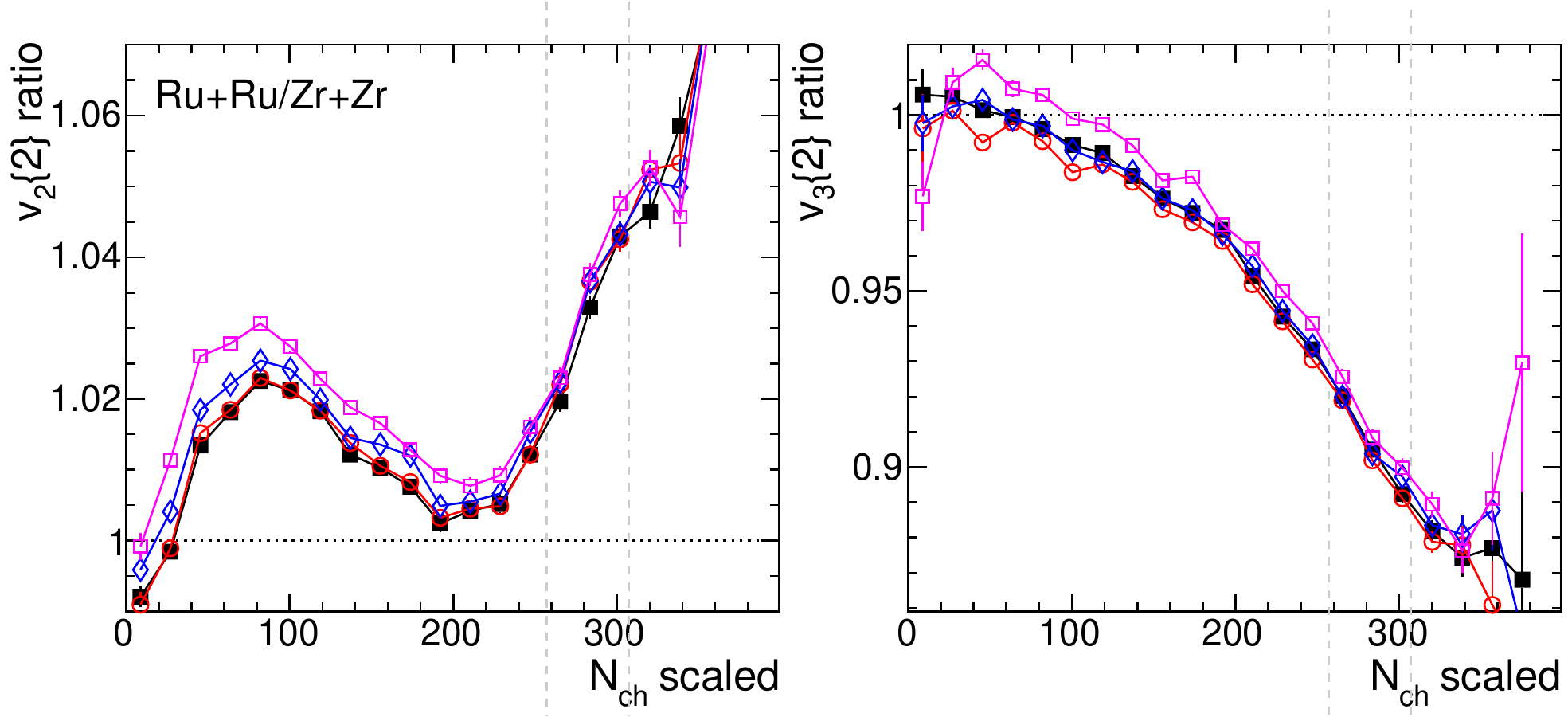}
\caption{\label{fig:8} The $v_2\{2\}$ (top left), $v_3\{2\}$ (top right) averaged over Ru+Ru and Zr+Zr collisions for four different $\eta$ ranges at the LHC energy. The bottom panels show the corresponding ratios between Ru+Ru and Zr+Zr collisions. }
\end{figure}

The last part of our analysis studies how the nuclear structure influences the rapidity dependence of flow observables. We perform this study at the LHC energy, and the results are presented in Fig.~\ref{fig:8}. The particles used to perform the flow analysis are taken from four $\eta$ ranges, while the $\nch$ used for the $x$-axis is always taken from $|\eta|<0.5$. 

We see that the $v_n\{2\}$ values decrease by about 20--30\% from $|\eta|<1$ to $3.5<|\eta|<5$. However, the isobar ratios change very little. The only noticeable difference is in the mid-central and peripheral collisions where ratios obtained in $3.5<|\eta|<5$ are slightly larger for both $v_2$ and $v_3$. We have traced this to a somewhat stronger response of $v_n$ to the $\Delta a_{0}$ between isobars. The observed similarity also suggests that the impact of longitudinal flow decorrelations is very small at LHC over the considered $\eta$ range. We conclude that the current implementation of the initial condition based on the HIJING model has too weak a longitudinal dependence in its response to the nuclear structure. Future model studies implementing realistic nPDF and the effects associated with dense gluon fields, such as gluon saturation physics, are necessary to quantify the sensitivity of the initial condition to nuclear structure effects more reliably.

\textit{\bf Summary} The ratios of the flow observables between $^{96}$Ru+$^{96}$Ru and $^{96}$Zr+$^{96}$Zr collisions are studied at $\snn=0.2$ and 5.02 TeV, and also as a function of pseudorapidity. These ratios are sensitive to structural differences between $^{96}$Ru and $^{96}$Zr nuclei regarding their nuclear shapes and radial profiles. The isobar ratios are similar between the two energies, though their response to structure difference is stronger at 5.02 TeV. The energy dependence is particularly noticeable in the impact of octupole deformation $\beta_3$ on the triangular flow $v_3$. 

We found that the effects of nuclear structure are captured entirely by the flow driven by the participant plane $v_{n,\mathrm{pp}}$, and are not carried by the difference between two-particle flow $v_n\{2\}$ and the $v_{n,\mathrm{pp}}$.  This observation suggests that the collective nuclear structure considered in this paper influences mainly the global geometry of the initial condition and, therefore, is captured by participant plane eccentricity $\varepsilon_n$. The isobar ratios have very weak pseudorapidity dependence, suggesting they are robust observables for detecting nuclear structure effects. Our study shows that the collision of isobar nuclei with different yet controlled structure differences at various beam energies, not specifically limited to $^{96}$Ru and $^{96}$Zr, can provide valuable information on the initial condition of heavy ion collisions. This study should be extended to other higher-order correlation observables, which are expected to exhibit stronger energy dependencies and better sensitivities to the initial condition~\cite{Bally:2022vgo}. 

In addition to the prospect of probing initial conditions using isobar collisions, our studies also reveal new insights generally valid in all large collision systems: We found that a large component of the $v_n$ is uncorrelated with the eccentricity but is generated dynamically during system evolution. Our results also suggest that $\varepsilon_n$ has multiple sources, and response coefficients for each source have different values and different energy dependencies. These new insights have significant impacts on the interpretation of the flow results obtained in RHIC and the LHC.

\textit{Acknowledgement.} We thank useful comments from Giuliano Giacalone. This work is supported by the U.S. Department of Energy under Grant No. DE-SC0024602.

\bibliography{../sqrtsandeta}{}
\bibliographystyle{apsrev4-1}
\end{document}